\documentclass[12pt]{iopart}

\usepackage[square,sort&compress,numbers]{natbib}
\usepackage{graphicx}
\usepackage{dcolumn}
\usepackage{bm}
\usepackage{subfig}
\usepackage{hyperref}
\usepackage[dvipsnames]{xcolor}

\begin{document}

\title{A transport approach to relate asymmetric protein segregation and population growth}

\author{Jiseon Min}
 \address{Department of Molecular and Cellular Biology, \\
 Harvard University, Cambridge, MA 02138, USA}
\author{Ariel Amir}%
 \ead{arielamir@seas.harvard.edu}
\address{%
John A. Paulson School of Engineering and Applied Sciences,
Harvard University, Cambridge, MA 02138, USA
}%

\date{\today}

\begin{abstract}
Many unicellular organisms allocate their key proteins asymmetrically between the mother and daughter cells, especially in a stressed environment. A recent theoretical model is able to predict when the asymmetry in segregation of key proteins enhances the population fitness, extrapolating the solution at two limits where the segregation is perfectly asymmetric (asymmetry $a$ = 1) and when the asymmetry is small ($0 \leq a \ll 1$) \cite{lin2019optimal}. We generalize the model by introducing stochasticity and use a transport equation to obtain a self-consistent equation for the population growth rate and the distribution of the amount of key proteins. We provide two ways of solving the self-consistent equation: numerically by updating the solution for the self-consistent equation iteratively and analytically by expanding moments of the distribution. With these more powerful tools, we can extend the previous model by \textit{Lin et al.} \cite{lin2019optimal} to include stochasticity to the segregation asymmetry. We show the stochastic model is equivalent to the deterministic one with a modified effective asymmetry parameter ($a_{\rm eff}$). We discuss the biological implication of our models and compare with other theoretical models.

\end{abstract}


%

\section{Introduction}
Genetically identical unicellular organisms may nevertheless exhibit different traits such as growth rate or generation time, a phenomenon known as phenotypic variability \cite{levien2020non}.
One cause of such diversity is the asymmetric allocation of key proteins at division. For instance, some bacteria and eukaryotic cells accumulate protein aggregates when exposed to stress and tend to segregate more of them to one of the two newborn cells at a division event. As a result, the ``rejuvenated'' cells (i.e. those with less damage) grow faster than the ``senescent'' cells \cite{Coelho2013, Coelho2015, Vaubourgeix2015, Vedel2016, Lindner2008, Stewart2005}. There have been a number of theoretical works \cite{Ackermann2007, lin2019optimal, Chao2010, Chao2016} trying to quantify the effect of asymmetric segregation of proteins to the overall growth of the population. In this paper, we strengthen the theory developed by \textit{Lin et al.} \cite{lin2019optimal} to relate the asymmetric protein segregation and the population growth rate by implementing a more general method involving transport equations and moment expansions. This allows us to recover previous results rigorously, and go beyond them. We extend the model to consider stochastic asymmetric segregation. Lastly, we compare alternative models of asymmetric segregation by applying the self-consistent equations.


\section{Model}

We define a generalization of the model proposed by \textit{Lin et al.} \cite{lin2019optimal}. Consider a clonal microbial population that grows exponentially with rate $\Lambda_p$. We assume that as a cell grows by volume $\Delta V$, the amount of key proteins $D$ increases by $\Delta D = S\Delta V$. Each cell's inter-division time ($\tau$) is a function of the initial concentration of the key protein ($\sigma_b$). For simplicity, suppose every cell is born with volume $V_b = 1$ and divides when the volume reaches $V_d = 2$. As it divides, the key proteins segregate asymmetrically according to the asymmetry parameter $a$: one daughter cell inherits $\frac{1+a}{2}$ of the key proteins from the mother and the other gets the rest. We call the key protein \emph{damaging} if the generation time $\tau[\sigma_b]$ is an increasing function of the initial protein concentration and \emph{beneficial} if it is a decreasing function.

\begin{figure}[!h]
\includegraphics[width = 15cm]{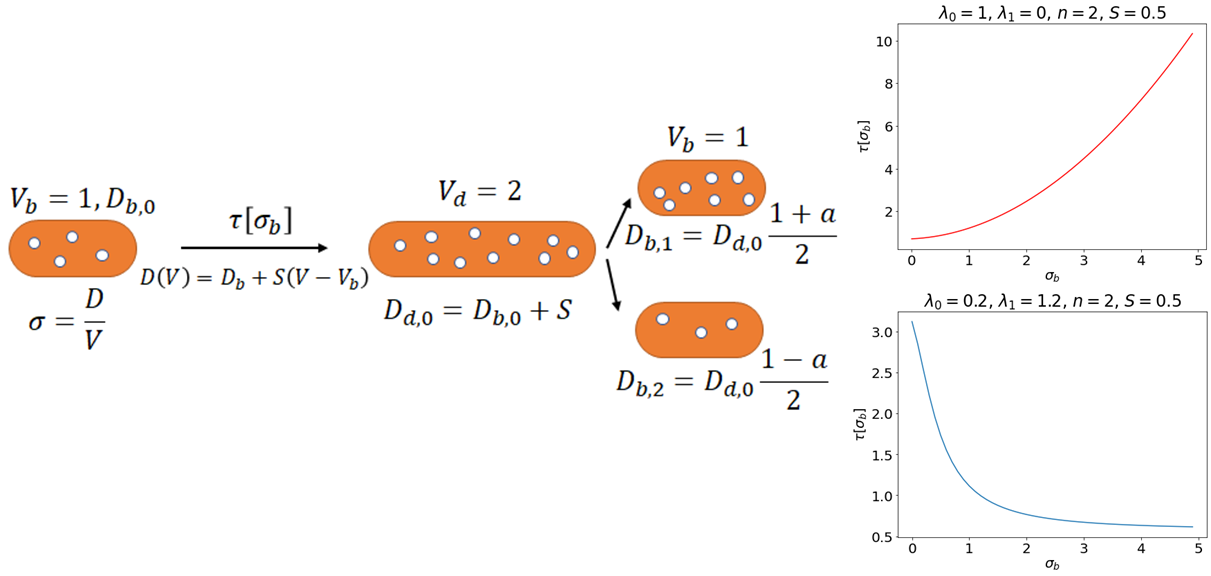}
\centering
\caption{Single cell level model. For simplicity, we assume that every cell has a volume $V_b = 1$ at birth and divides when it reaches $V_d = 2$. The amount of key proteins in a cell increases in proportion to the added volume with a prefactor $S$. The generation time $\tau[\sigma_b]$ is a function of the key protein concentration at birth ($\sigma_b = D_b / V_b$). Asymmetry $a \in [0, 1]$ tunes the ratio of key proteins inherited by two offspring cells ($a = 0$ being perfectly asymmetric, $a = 1$ being perfectly asymmetric). The plots on the right show an example of generation time as a function of initial concentration of key protein which is deleterious (top) or beneficial (bottom).}
\end{figure}

While our analysis can be applied to any analytical function $\tau[\sigma_b]$, we assume the instantaneous growth rate $\lambda[\sigma]$ is a Hill-type function of the key protein concentration of the cell, as proposed by \textit{Lin et al.} \cite{lin2019optimal}. This assumptions by \textit{Lin et al.} is based on observations made in experimental studies, where growth rate function of microbes often has an inflection point \cite{Lindner2008, Vedel2016}.

\begin{eqnarray}
	\lambda[\sigma] &= \frac{\lambda_0 + \lambda_1 \sigma^n}{1 + \sigma^n},\\
	\tau[\sigma_b] &= \int_1^2 \frac{1}{V \lambda[\sigma]} dV\nonumber\\
		&= \int_1^2 \frac{1}{V \lambda[(\sigma_b - S) / V + S]} dV. \label{physiological}
\end{eqnarray} 

Every time a cell divides, the asymmetry parameter $a$ is randomly and independently drawn from a probability distribution $g(a)$. The deterministic model by \textit{Lin et al.} \cite{lin2019optimal} assumes a delta function for $g(a)$, but here we allow $a$ to fluctuate.

\section{Results}
\subsection{Transport equation}
Recently, \textit{Levien et al.} \cite{levien2020interplay} showed that a transport equation can be used to get a self-consistent equation for the population distribution of phenotype and population growth rate even when the properties of the cells are correlated across the generations. Here we follow their derivation by setting the phenotype of interest as the key protein concentration at birth $\sigma_b$.

Assume that the population size is $N$ at time $t$ and it grows exponentially with the rate $\Lambda_p$. Let us also assume that the population has reached a stable probability distribution $\psi(u, \sigma_b, t)$, where $u$ is time since birth, $\sigma_b$ is the key protein concentration at birth, and $t$ is the global time. The population then evolves during an infinitesimal time step $dt$ as:
\begin{eqnarray}
	N\psi(u, \sigma_b, t) &= N e^{-\Lambda_p dt} \psi(u - dt, \sigma_b, t - dt), \nonumber \\
	&= N (1 - \Lambda_p dt) \left(\psi(u, \sigma_b, t) - dt \frac{\partial \psi}{\partial u} - dt \frac{\partial \psi}{\partial t}\right).
\end{eqnarray}

Rearranging the terms and dividing both sides by $dt$, we get
\begin{eqnarray}
	\frac{\partial \psi}{\partial u} + \frac{\partial \psi}{\partial t} = - \Lambda_p \psi.
\end{eqnarray}

In a steady state population, there is no dependence on $t$, so $\partial_t \psi = 0$. Therefore, 
\begin{eqnarray}
	\frac{\partial \psi(u, \sigma_b)}{\partial u} &= - \Lambda_p \psi(u, \sigma_b),\\
	\psi(u, \sigma_b) &= \psi(0, \sigma_b) e^{-\Lambda_p u}.\label{sol}
\end{eqnarray}

The boundary conditions are set by the division. The probability density of having a newborn cell with a protein concentration $\sigma_b$ is 
\begin{eqnarray}
	\psi(0, \sigma_b) &= 2 \int \int f(\sigma_b|\sigma_b', a') g(a') \psi(\tau[\sigma_b'], \sigma_b') d\sigma_b' da'.
\end{eqnarray}
The probability density $f(\sigma_b|\sigma_b', a')$ and $g(a')$ depend on the physiological model we choose. For instance, in a deterministic model, we have
\begin{eqnarray}
	&&f(\sigma_b|\sigma_b', a') = \frac{1}{2} \delta \left(\sigma_b - \frac{1+a'}{2} (\sigma_b' + S)\right) + \frac{1}{2} \delta \left(\sigma_b' - \frac{1-a'}{2} (\sigma_b' + S),\right) \label{f}\\
	&&g(a') = \delta(a' - \bar{a}). \label{g}
\end{eqnarray} 

Applying the solution given by Equation \ref{sol} (in the general case), we can eliminate the dependence on $u$. Thus, we can rewrite to obtain a self-consistent equation:
\begin{eqnarray}
	\psi_{\rm birth}[\sigma_b] &= 2 \int \int f(\sigma_b| \sigma_b', a') \psi_{\rm birth}[\sigma_b'] g(a') e^{-\Lambda_p \tau[\sigma_b']} d\sigma_b' da'. \label{self_consistent}
\end{eqnarray}
In the next section, we apply the self-consistent Equation \ref{self_consistent} to a toy model of \textit{E. coli} recently published by \textit{Blitvic et al.} \cite{blitvic2020aging} as a simple example. 

\subsection{Toy model of the aging of \textit{E. coli} \cite{blitvic2020aging} }

Let us introduce a simple model published by \textit{Blitvic et al.} to demonstrate how to use our approach to connect the asymmetry between two newborn cells and population growth rate \cite{blitvic2020aging}. \textit{Blitvic et al. }assume that as an \textit{E. coli} cell divides, one daughter cell is rejuvenated and the other senescent where the generation time of a rejuvenated cell ($\tau[R]$) is shorter than a senescent cell ($\tau[S]$). In this model, there are two phenotype $x = S$ for a senescent cell and $x = R$ for a rejuvenated one.

Then the transition density $f(x|x')$ from a phenotype $x'$ to $x$ is
\begin{eqnarray}
	f(x|x') &= \frac{1}{2} \delta(x - R) + \frac{1}{2} \delta(x - S).
\end{eqnarray}

The self-consistent equation is
\begin{eqnarray}
	\psi[x] &= \int \psi[x'] f(x|x') e^{-\Lambda_p \tau[x']} dx'.
\end{eqnarray}
Integrating both sides with respect to $x$, we have
\begin{eqnarray} \label{hartwell}
	1 &= e^{-\Lambda_p \tau[R]} + e^{-\Lambda_p \tau[S]}.
\end{eqnarray}

Because the exponential function is convex, by Jensen's inequality,
\begin{eqnarray}
	1 = e^{-\Lambda_p \tau[S]} + e^{-\Lambda_p \tau[R]}> 2 e^{-\Lambda_p \left<\tau[x]\right>},
\end{eqnarray}

where $\left<\tau[x]\right> = \frac{1}{2} (\tau[S] + \tau[R])$. Thus, the asymmetry of generation time of two distinct phenotypes enhances the population growth rate if the mean of the generation time, $\left<\tau[x]\right>$ is kept constant. Note that Equation \ref{hartwell} was also derived in Ref. \cite{hartwell1977unequal} in the context of the asymmetric division of the budding yeast \textit{Saccharomyces cerevisiae}.

In the next section, we solve Equation \ref{self_consistent} numerically and analytically more generally. 

\subsection{\label{iterate} Numerically solving the self-consistent equation}

The self-consistent equation \ref{self_consistent} can be numerically solved by iterating the following procedure. First, we start with initial guesses for the solutions $\psi_{\rm birth}^{(0)}[\sigma_b]$ and $\Lambda_p^{(0)}$. We then update our guess for the distribution $\psi_{\rm birth}[\sigma_b]$ using Equation \ref{self_consistent}. Next, we update $\Lambda_p$ by using the Euler-Lotka equation \cite{levien2020interplay}, obtained by integrating Equation \ref{self_consistent} with respect to $\sigma_b$. This procedure is summarized in this set of equations:

\begin{eqnarray}
	&&\widetilde{\psi}_{\rm birth}^{(n)}[\sigma_b] = 2 \int f(\sigma_b| \sigma_b', a')\psi_{\rm birth}^{(n-1)}[\sigma_b'] g(a') e^{-\Lambda_p^{(n-1)} \tau[\sigma_b']} da' d\sigma_b', \label{it1}\\
	&&\psi_{\rm birth}^{(n)}[\sigma_b] = \frac{\widetilde{\psi}_{\rm birth}^{(n)}[\sigma_b]}{\int d\sigma_b' \widetilde{\psi}_{\rm birth}^{(n)}[\sigma_b']} \label{it2}\\ 
	&&1 = 2 \int \int \psi_{\rm birth}^{(n)}[\sigma_b'] g(a') e^{-\Lambda_p^{(n)} \tau[\sigma_b']} da' d\sigma_b'. \label{it3}
\end{eqnarray}

$\psi_{\rm birth}^{(n)}[\sigma_b]$ and $\Lambda_p^{(n)}$ converge to constant values as $n \rightarrow \infty$. Note that the solution is not sensitive to the choice of $\psi_{\rm birth}^{(0)}$ and $\Lambda_p^{(0)}$. In \ref{distribution}, we discuss the shape of the distribution $\psi_{\rm birth}[\sigma_b]$.

\begin{figure}[!h]
\centering
\subfloat[]{{\includegraphics[width = 7.5cm]{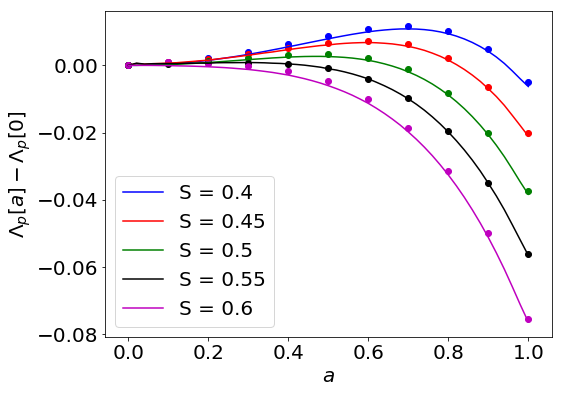} }}
\qquad
\subfloat[]{{\includegraphics[width = 7.5cm]{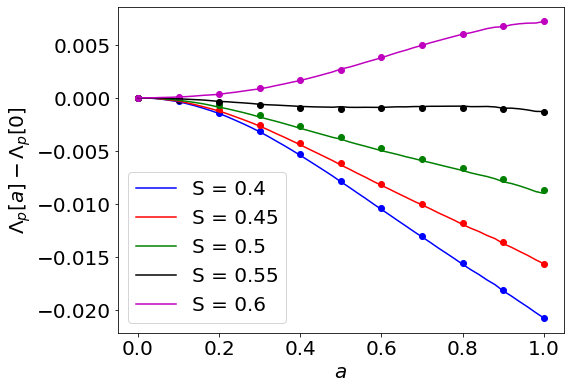} }}
\caption{The relative population growth rate, compared to $a = 0$, is plotted against the asymmetry $a$. The solid lines show the numerical solutions of Equation \ref{self_consistent} and the circles mark simulated population growth rates. (a) Beneficial protein ($\lambda_0 = 0.2, \lambda_1 = 1.2, n = 2$), (b) Deleterious protein ($\lambda_0 = 1, \lambda_1 = 0, n = 2$). For numerical solutions, we use $\psi_{\rm birth}^{(0)}[\sigma] = \delta(\sigma_b - S)$ and $\Lambda_p^{(0)} = \ln 2/\lambda[S]$. Error bars are smaller than the marker size.}
\label{PDE_plots}
\end{figure}

In Figure \ref{PDE_plots}, we compare the numerical solution for Equation \ref{self_consistent} in the case of deterministic $a$ (i.e. $g(a')$ is a delta function) and the fitted exponential growth rate obtained by simulating the population directly (see \ref{simulation} for details of the simulation).

Figure \ref{PDE_plots} confirms that Equation \ref{self_consistent} has sufficient information for calculating the population growth rate. The iterative procedure is more accurate, faster, and less memory-intensive than a brute force simulation which involves exponentially growing population size. However, the main shortcoming of the numerical solution is that it does not give any insight about how $\Lambda_p$ depends on $S$ and $a$. It is possible infer the relation by finding $\Lambda_p$ for various values of $S$ and $a$, but for every parameter values, we have to go through Equations \ref{it1} - \ref{it3}, which is highly inefficient. For those reasons, the moment expansion method we discuss in the following section is a good alternative way of calculating $\Lambda_p$ for small $a$.

\subsection{Moment expansion \label{momentexpansion}}
Equation~\ref{self_consistent} can be approximately solved by expanding the moments of $\psi_{\rm birth}[\sigma_b]$. For simplicity, we start with the deterministic case. As we will explain in the following sections, it is straightforward to generalize this moment expansion method to the stochastic case.

When the key protein segregation is perfectly symmetric (i.e. $a = 0$), every cell has a concentration of key protein of $\sigma_b = S$ at cell birth. Therefore, for a small $a$, the distribution $\psi_{\rm birth}[\sigma_b]$ should be sharply peaked around $S$. Based on this, we are motivated to expand the moment of the distribution around $\sigma_b = S$ by defining $\Delta \sigma_b = \sigma_b - S$. In order to calculate the $n$-th moment, we multiply both sides of Equation \ref{self_consistent} by $\Delta \sigma_b^n$ and integrate them with respect to $\sigma_b$. We use Equations \ref{f} and \ref{g} to specify $f(\sigma_b|\sigma_b', a')$ and $g(a')$. Invoking Newton's binomial theorem, we obtain:
\begin{eqnarray}
	\langle \Delta \sigma_b^n \rangle &= \frac{1}{2^{n-1}} \sum_{m = 0}^{\lfloor n/2\rfloor} {n \choose 2m}  a^{2m}\left<\Delta \sigma_b^{n - 2m} (\Delta \sigma_b + 2 S)^{2m} e^{-\Lambda_p \tau[\Delta \sigma_b + S]} \right>.  \label{moment} 
\end{eqnarray}

Note that by convention, ${n \choose 0} = 1$ and $\lfloor n/2 \rfloor$ is the greatest integer that is less than or equal to $n/2$. From the right hand side of Equation~\ref{moment}, we can infer that $\langle \Delta \sigma_b^n\rangle = O(a^{2 \lfloor n/2\rfloor})$. This means that by solving a system of $2k + 1$ equations ($n = 0, \cdots, 2k$), we will find $\Lambda_p$ to $O(a^{2k})$ accuracy. For each equation, we need to expand the exponential function $e^{-\Lambda_p \tau[\Delta \sigma_b +S]}$ around $\Delta \sigma_b = 0$ to order $a^{2\lfloor n/2\rfloor}$. To illustrate the process more clearly, let us calculate $\Lambda_p$ to $O(a^2)$ accuracy.

If $a = 0$, every cell has $\sigma_b = S$ when the population is stabilized, so $\Lambda_p = \frac{\ln 2}{\tau[S]}$. Due to symmetry, $\Lambda_p[a] = \Lambda_p[-a]$. Thus, to solve $\Lambda_p$ to $O(a^2)$ is equivalent to finding a coefficient $C_2$ defined as below,
 
\begin{eqnarray}
	\Lambda_p = \frac{\ln2}{\tau[S]} (1 + C_2 a^2 + O(a^4)).
\end{eqnarray}

Taylor expanding $\tau[\Delta \sigma_b + S]$ around $\Delta \sigma_b = 0$ to $O(a^2)$ we find
\begin{eqnarray}
	\tau[\Delta \sigma_b + S] &= \tau[S] + \tau'[S] \Delta \sigma_b + \frac{\tau''[S]}{2} \Delta \sigma_b^2.
\end{eqnarray}

Expanding $e^{-\Lambda_p \tau[\Delta \sigma_b + S]}$ to $O(a^2)$ leads to:
\begin{equation}
\hspace*{-2cm}
	e^{-\Lambda_p \tau[\Delta \sigma_b + S]} = \frac{1}{2} \left[1 - \ln 2 \left( C_2 a^2  + \frac{\tau'[S]}{\tau[S]}\Delta \sigma_b + \left(\frac{\tau''[S]}{2\tau[S]} - \frac{\ln 2}{2} \frac{\tau'[S]^2}{\tau[S]^2}\right) \Delta \sigma_b^2 \right) \right].
\end{equation}
Applying these to Equation \ref{moment} while setting $n =0, 1, $ or $2$, we can set up a system of three linear equations with the unknowns being $C_2, \left<\Delta \sigma_b\right>$ and $\left<\Delta \sigma_b^2\right>$. (Note that here $\tau[S]$ is a known function specified in the model.)

\begin{eqnarray}
	 &1 = 1 - \ln 2 \left[C_2  a^2  + \frac{\tau'[S]}{\tau[S]} \langle \Delta \sigma_b \rangle + \frac{\tau''[S]}{2\tau[S]}\langle \Delta \sigma_b^2 \rangle  - \frac{\ln 2}{2}\frac{\tau'[S]^2}{\tau[S]^2}\langle \Delta \sigma_b^2 \rangle \right], \\
	 &\langle \Delta \sigma_b \rangle = \frac{\langle \Delta \sigma_b \rangle}{2} - \frac{\ln 2}{2} \frac{\tau'[S]}{\tau[S]} \langle \Delta \sigma_b^2\rangle,\\
	 &\langle \Delta \sigma_b^2 \rangle = \frac{\langle \Delta \sigma_b^2\rangle}{4} +S^2 a^2. 
\end{eqnarray}

Solving these, we get
\begin{eqnarray}
	\langle \Delta \sigma_b \rangle &= - \frac{4 \ln 2}{3} \frac{\tau'[S]}{\tau[S]}S^2 a^2, \\
	\langle \Delta \sigma_b^2 \rangle &= \frac{4}{3} S^2 a^2,\\
	C_2 &= -\frac{2}{3} \frac{\tau''[S]}{\tau[S]} S^2 + 2 \ln 2 \frac{\tau'[S]^2}{\tau^2[S]}S^2. 
\end{eqnarray}

Therefore, the population growth rate is 
\begin{eqnarray}
	\Lambda_p &= \frac{\ln2}{\tau[S]} \left(1 + \left(-\frac{2}{3} \frac{\tau''[S]}{\tau[S]} + 2 \ln 2 \frac{\tau'[S]^2}{\tau^2[S]}\right)S^2a^2 \right).
\end{eqnarray}

This is general to any choice of $\tau[\sigma_b]$, but in particular, if the generation time is described by Equation~\ref{physiological}, we can rewrite the population growth rate in terms of $\lambda[S]$.

\begin{eqnarray}\label{a_near_0_sol}
	\Lambda_p &= \lambda[S] \left(1 + \frac{\lambda''[S]}{4 \ln 2 \lambda[S]} S^2 a^2 \right).
\end{eqnarray}

This second order solution for $\Lambda_p$ is consistent with the result by \textit{Lin et al. }\cite{lin2019optimal} but is derived here more rigorously and more generally since it applies to any functional form of $\tau[\sigma_b]$. Furthermore, we can calculate $\Lambda_p$ to an arbitrarily high order as desired by solving for $\Lambda_p = \frac{\ln 2}{\tau[S]} \left(1 + C_2 a^2 + C_4 a^4 + \cdots\right)$, as we elaborate in \ref{higher order}. 

\section{Phase transition of $a_c$}

In the previous sections, we explained how to solve Equation \ref{self_consistent} and predict the population growth rate $\Lambda_p$ as a function of growth rate of a single cell $\lambda[\sigma]$, protein accumulation rate $S$, and asymmetry $a$. In the context of evolution, we can imagine that a single-celled organism may adapt to environmental stress (characterized by $S$ and $\lambda[\sigma]$ in our model) by tuning the asymmetry $a$. In population genetics, fitness of a population is often defined as an exponential growth rate of the population size, and traits that increase the fitness tend to survive longer due to natural selection \cite{hartl1997principles}. In fact, there are several experimental studies reporting that segregation of protein aggregates tend to become more asymmetric as temperature increases above the optimal temperature \cite{Coelho2013, Coelho2015, Vaubourgeix2015, Vedel2016}. In this section, we present how to find the optimal asymmetry $a_c$, the value of $a$ that maximizes the population growth rate $\Lambda_p$, as a function of population accumulation rate $S$. As \textit{Lin et al. }\cite{lin2019optimal} did, we find a sharp phase transition of $a_c$ if the key protein is deleterious and a smooth transition in the case of beneficial key protein.

Previously, \textit{Lin et al.} \cite{lin2019optimal} inferred $a_c[S]$ by interpolating the solutions near $a = 0$ and at $a = 1$ with a minimal number of extrema. \textit{Lin et al.} used a Landau approach to find the population growth rate $\Lambda_p$ to $O(a^2)$ accuracy for a small $a$. In summary, the Landau approach investigates the shape of the fitness curve $f = \Lambda_p[a] - \Lambda_p[a = 0]$ as a function of $a$ at a given key protein accumulation rate $S$ (i.e. the fitness is analogous to a free energy with an order parameter $a$, and $S$ to temperature). They derived the relationship between the fitness $f$ and asymmetry $a$ from an equation for the total cell volume, $V = \sum_i v_i(t)$. 

\begin{equation}
	\frac{dV}{dt} = \sum_i \lambda[\sigma_{b, i}] v_i.
\end{equation}

Then they Taylor expand $\lambda[\sigma]$ around $\sigma = S$.

\begin{equation}
	\frac{dV}{dt} = V\lambda[S] + \sum_i v_i \left( \left. \frac{d\lambda}{d\sigma}\right\vert_S(\sigma_{b,i} - S) + \frac{1}{2}\left. \frac{d^2\lambda}{d\sigma^2}\right\vert_S(\sigma_{b,i} - S)^2 + \cdots \right).
\end{equation}

Using the fact that odd power terms of $a$ should vanish due to symmetric cell division in terms of volume and an ansatz that the variance of the distribution of $\sigma_b$ scales as $a^2$, \textit{Lin et al.} derived Equation \ref{a_near_0_sol} without any knowledge about the distribution $\psi[\sigma_b]$.

\begin{figure}[!h]
\includegraphics[width = 7cm]{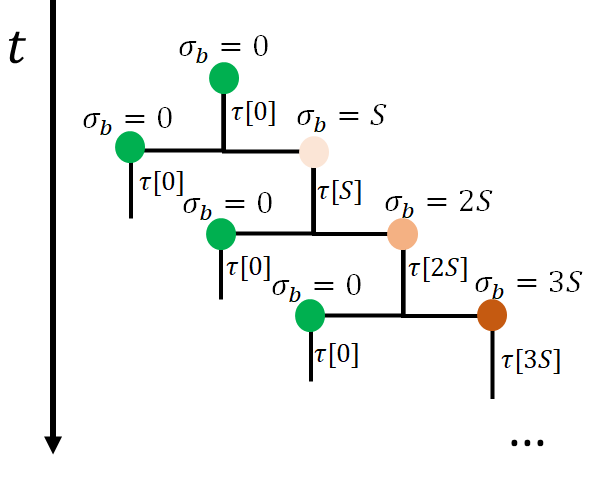}
\centering
\caption{The lineage tree for a population with $a = 1$ is self-similar. Specifically, a population starting with a cell with no key protein at birth ($\sigma_b = 0$, marked by a green circle) should all have the same population size as a function of time since birth of the ancestor. Using this property \textit{Lin et al.} derived Equation \ref{lineqn} \cite{lin2019optimal}.}
\label{self_similar_tree}
\end{figure}

Equation \ref{a_near_0_sol} breaks down as $a$ approaches 1, but when $a$ is exactly 1, there exists a closed formula for $\Lambda_p$ due to the self-similarity in a population tree as illustrated in Figure \ref{self_similar_tree}. \textit{Lin et al.} showed that 

\begin{equation}\label{lineqn}
	\sum_{i = 0}^\infty \exp\left( - \Lambda_p[a = 1] \sum_{j = 0}^i \tau[jS] \right) = 1.
\end{equation}

\begin{figure}[!h]
\includegraphics[width = 15cm]{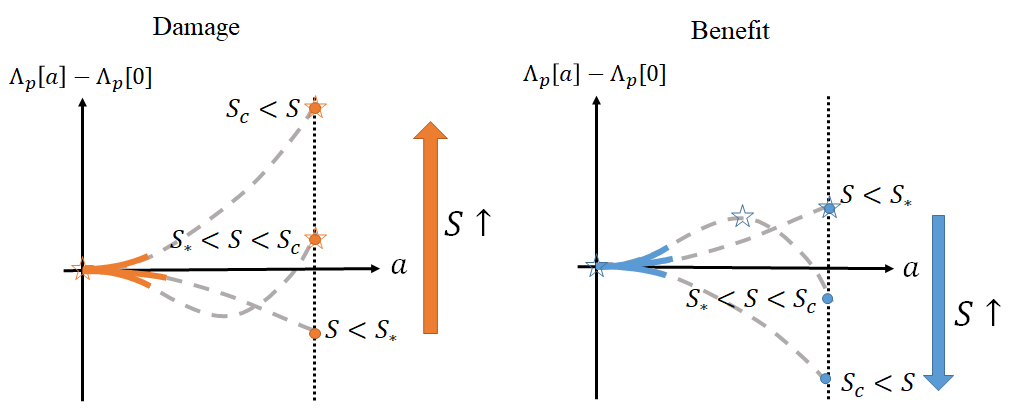}
\centering
\caption{Finding the optimal asymmetry $a_c$ by graphical interpolation \cite{lin2019optimal}. \textit{Lin et al.} find the second derivative of the fitness ($\Lambda_p[a] - \Lambda_p[0]$) near $a = 0$ (solid line) and numerically solve the self-consistent equation for $a = 1$ (Equation \ref{lineqn}, filled circles). Interpolating between the two limits with a minimal number of extrema, the authors find what value of $a$ maximizes the fitness (empty stars). Unlike in the damage case, in the benefit case, there is a large range of $S$ where the fitness curve is convex near $a = 0$ and $\Lambda_p[a = 1] < \Lambda_p[a=0]$ (i.e. $S_* < S < S_c$). This means that the population growth rate is maximized when $a$ is between 0 and 1. }
\label{interpolation_figure}
\end{figure}

Interpolating the solutions at $a \approx 0$ and $a = 1$ with the minimal number of extrema, \textit{Lin et al.} infers the value of $a_c$ (i.e. what value of $a$ maximizes $\Lambda_p$ for a given $S$) as described in Figure \ref{interpolation_figure}. They found that the location of $a_c$ depends on the relative magnitude of two critical values of $S$: $S_c$ and $S_*$. Each of these two critical values arises from the solutions at $a \approx 0$ and $a = 1$. $S_c$ is where the second derivative of $\lambda[S]$ is zero, which means that the fitness near $a = 0$ changes its sign when $S = S_c$. $S_*$ is where $\Lambda_p[a = 1] = \Lambda_p[a = 0]$. See Figure \ref{interpolation_figure} for the inferred shape of the fitness curve at various values of $S$ relative to $S_c$ and $S_*$. Note that $a_c$ is between 0 and 1 (marked by star in Figure \ref{interpolation_figure}) in the case of beneficial key protein when $S$ is between $S_*$ and $S_c$.





Similar to the previous paper by \textit{Lin et al.}, we use an interpolation to infer where the phase transition of $a_c$ is. In addition, we are equipped with a more rigorous way of finding the population growth rate near $a = 0$ as described in Section \ref{momentexpansion}. We also have a transcendental equation about the slope of $\Lambda_p[a]$ at $a = 1$, which we discuss in depth in \ref{a_near_1}. We denote the solution to Equation \ref{lineqn} as $\Lambda_{p1}$ and expand $\Lambda_p[a] = \Lambda_{p1} (1 + C_1 (1 - a) + O((1 - a)^2))$. Finding the sign of $C_1$ makes it possible to predict whether $\Lambda_p[a = 1]$ is a local minimum $(C_1 > 0)$ or a local maximum $(C_1 < 0)$. Using a similar moment expansion approach as we do for $a \approx 0$, we find the first coefficient $C_1$ in \ref{a_near_1} as

\begin{equation}
\hspace*{-2cm}
	C_1 = \frac{S}{2} \frac{\sum_{i = 0 }^\infty e^{-\Lambda_{p1} \sum_{j = 0}^i \tau[jS]} \sum_{k = 0}^i \tau'[kS] \left(\sum_{m = 0}^\infty (m + 1)e^{-\Lambda_p \sum_{n = 0}^m \tau[nS]} - \frac{k (k + 1)}{2}\right)}{\sum_{i = 0}^\infty e^{-\Lambda_{p1} \sum_{j = 0}^i \tau[jS]} \sum_{k = 0}^i \tau[kS]}
\end{equation}

\begin{figure}[!h]
\includegraphics[width = 9.5cm]{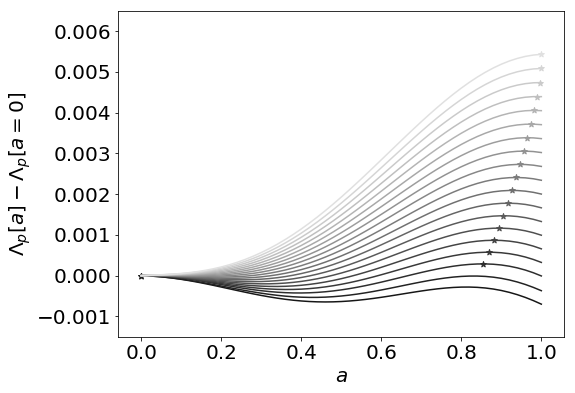}
\centering
\caption{Increase of $\Lambda_p$ relative to $a = 0$ against $a_c$ calculated by numerically solving Equation \ref{self_consistent}. $S$ increases as the line gets lighter ($S = 0.554$ to $0.592$). The stars mark $a_c$ for each $S$.  If $ S < S_*$, then automatically, $S < S_c$ and $S < S_c^{a = 1}$. This means that $a = 0$ is a local maximum and there is also another peak between $a = 0$ and $a = 1$. At $S = S_*$, $S < S_c^{a = 1}$, which means that there is $a <1$ that has a higher $\Lambda_p$ than $a = 0$ and $a = 1$. Therefore, there should be a first-order phase transition at $S < S_*$. }
\label{zoom_in}
\end{figure}

Here we can define another critical value $S_c^{a = 1}$ which is the value of $S$ that makes $C_1 = 0$. Because $\tau'[kS] > 0(\tau'[kS] < 0)$ for the damage (benefit) case, $C_1$ starts out positive (negative) for a small $S$ and becomes negative (positive) as $S$ exceeds $S_c^{a = 1}$. In other words, when $S < S_c^{a = 1}$, $a = 1$ is a local minimum (maximum) of $\Lambda_p[a]$, whereas it becomes a local maximum (minimum) when $S > S_c^{a = 1}$ in the damage (benefit) case. This provides us an additional piece of information that we can use to infer whether $a_c = 1$ for a given $S$. For instance, in the case of damage ($\lambda_0 = 1, \lambda_1 = 0$), $S_c^{a = 1} \approx 0.592$ is greater than $S_* \approx 0.558$. This means that if $S$ is between $S_*$ and $S_c^{a = 1}$, $\Lambda_p[a]$ should have a local maximum below $a = 1$. When $S$ is above both $S_*$ and $S_c^{a = 1}$, the population growth rate is maximized at $a = 1$ (i.e. $a_c = 1$). This means that the transition of $a_c$ is not a single discrete jump from $a_c = 0$ to $a_c = 1$, but rather a jump to an intermediate $a_c \in (0, 1)$ and then a smooth increase to $a_c = 1$. This process is illustrated in Figure \ref{zoom_in}. 

\begin{figure}[!h]
\centering
\subfloat[]{{\includegraphics[width = 7.5cm]{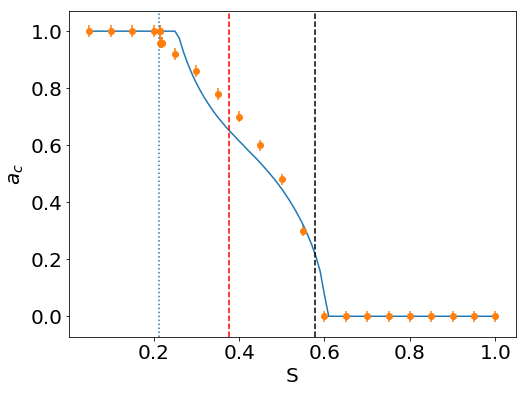} }}
\qquad
\subfloat[]{{\includegraphics[width = 7.5cm]{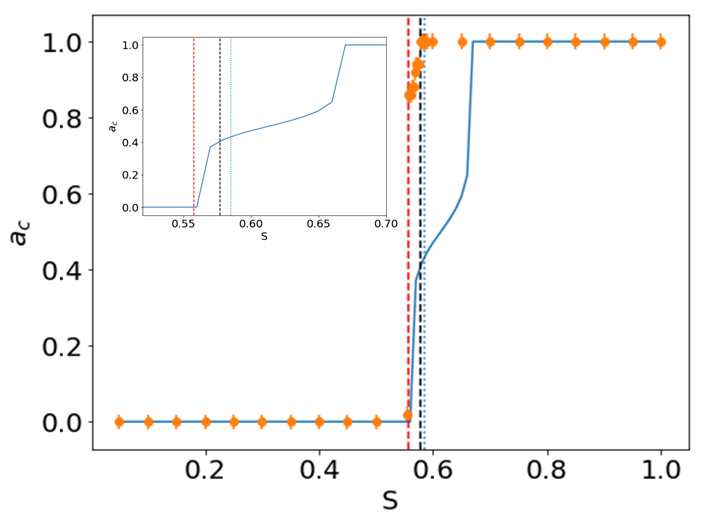} }}
\caption{$a_c$, the value of $a$ that maximizes the population growth rate, plotted against $S$. The solid blue lines are the prediction made from the sixth order solution of $\Lambda_p$. The orange points, each of which has an error bar, show $a_c$ found by solving Equation \ref{self_consistent} numerically as described in section \ref{iterate}. (a) Beneficial protein ($\lambda_0 = 0.2, \lambda_1 = 1.2, n = 2$), (b) Deleterious protein ($\lambda_0 = 1, \lambda_1 = 0, n = 2$). The inset zooms in on the region around $S=0.6$. The dotted blue line shows where the slope of $\Lambda_p$ at $a = 1$ changes its sign. The red dashed line is at $S = S_*$ where $\Lambda_p[a = 0] = \Lambda_p[a=1]$ and the black dashed line is at $S = S_c$ where $\lambda''[S] = 0$ (i.e. the second derivative of $\Lambda_p[a]$ changes its sign at $a = 0$). }
\label{ac_plots}
\end{figure}

In the case of beneficial key proteins, \textit{Lin et al.} found that $a_c = 1$ for small $S$ and decreases smoothly to $a_c = 0$ as $S$ increases, but the start of that transition was not specified. Based on \ref{a_near_1}, we find that $a_c$ starts deviating from one at $S_c^{a=1} = 0.214$ (blue dotted line in Figure ~\ref{ac_plots} (a)), consistent with the numerical solution of Equation \ref{self_consistent}. Comparing the solid blue line ($a_c$ predicted by the moment expansion to the sixth order) and the orange dots (numerical simulation), note that the moment expansion method does not successfully predict $a_c$ when the true $a_c$ value is close to $1$. However, as $S$ increases, and $a_c$ becomes closer to $0$, the answers agree better, since $a^2 \approx 0 $ our approximation holds. 

\section{Introducing noise}

The moment expansion method can be easily generalized to the stochastic model where $a$ is drawn from a distribution $g(a)$ every time a cell divides. Equation~\ref{moment} becomes

\begin{equation}
\hspace*{-2cm}
	\langle \Delta \sigma_b^n \rangle = \frac{1}{2^{n-1}} \sum_{m = 0}^{\lfloor n/2\rfloor} {n \choose 2m}  \int g(a) a^{2m} da \left<\Delta \sigma_b^{n - 2m} (\Delta \sigma_b + 2 S)^{2m} e^{-\Lambda_p \tau[\Delta \sigma_b + S]} \right>.
\end{equation}

This implies that in the small $a$ limit, we can define an effective $a$, $a_{\rm eff} = \sqrt{\langle a^2\rangle}$, for which the population growth rate $\Lambda_p$ has the same formula independent of $g(a)$.
\begin{eqnarray}
	\Lambda_p &= \lambda[S] \left(1 + \frac{\lambda''[S]}{4 \ln 2 \lambda[S]} S^2 a_{\rm eff}^2 \right).
\end{eqnarray}

\begin{figure}[!h]
\centering
\subfloat[]{{\includegraphics[width = 7.5cm]{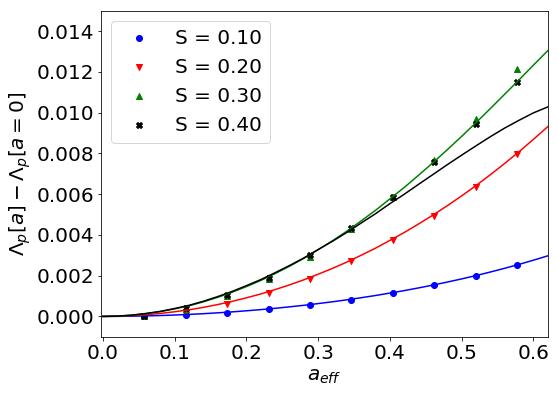} }}
\qquad
\subfloat[]{{\includegraphics[width = 7.5cm]{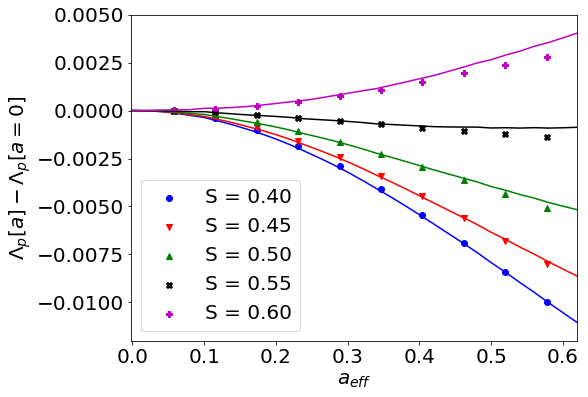} }}
\caption{The change of population growth rate ($\Lambda_p[a_{\rm eff}] - \Lambda_p[a = 0]$) plotted against the effective value of $a$ where $g(a) = U(\bar{a} - \sigma_a, \bar{a} + \sigma_a)$, a uniform distribution. The solid lines show the numerical solutions of Equation \ref{self_consistent} for $\sigma_a = 0$ and the circles correspond to simulated values where $\bar{a} = 0$ and $\sigma_a = 0, 0.1, 0.2, \cdots, 1$. (a) Beneficial protein ($\lambda_0 = 0.2, \lambda_1 = 1.2, n = 2$), (b) Deleterious protein ($\lambda_0 = 1, \lambda_1 = 0, n = 2$).}
\label{siga}
\end{figure}

In Figure~\ref{siga}, we numerically solve Equation \ref{self_consistent} where $g(a)$ is a uniform distribution ($g(a) = U(\bar{a} - \sigma_a, \bar{a} + \sigma_b)$ with $\bar{a} = 0$ and $\sigma_a =0, 0.1, 0.2, \cdots 1$) and plot the population growth rate against $a_{\rm eff} = \sqrt{\left< a^2 \right>}$. As predicted, $a_{\rm eff}$ behaves exactly as $a$ does in the deterministic model for a broad range of $\sigma_a$. For a relatively large $S$ and large $\sigma_a$, the population growth rate obtained by simulating the exponentially growing population starts to deviate from the second order prediction as shown in Figure~\ref{siga}. This conclusion that stochastic segregation of key proteins increases effective asymmetry is analogous to the finding made by \textit{Chao et al.} \cite{Chao2016} about damage partitioning in \textit{E. coli} that larger fluctuations result in similar increase of the population growth as a fixed asymmetric ratio.


\section{Asymmetry $a$ as a variable of concentration of key proteins}
So far, we have considered a model where the segregation ratio of key proteins is independent from the amount of key proteins each cell has. However, there is experimental evidence that the degree of asymmetry may depend on how much key proteins a cell has \cite{Coelho2013, Stewart2005}. \textit{Vedel et al.} \cite{Vedel2016} studied how protein aggregates in \textit{E. coli} created in responds to heatshock are segregated differently between two daughters. One of their results is that the asymmetric segregation parameter $a$ is a function of the amount of protein aggregates $D$, and if $a$ is assumed to be constant, the fitness is underestimated. We can test the effect of dependence of $a$ to $D$ by replacing $a$ with $a[D]$ in Equation \ref{self_consistent}.

\begin{eqnarray}
	\psi_{\rm birth}[\sigma_b] &= 2\int \psi_{\rm birth}[\sigma_b'] f(\sigma_b|\sigma_b', a[\sigma_b' + S]) e^{-\Lambda_p \tau[\sigma_b']} d\sigma_b'.
\end{eqnarray}

\begin{figure}[!h]
\centering
\includegraphics[width = 10cm]{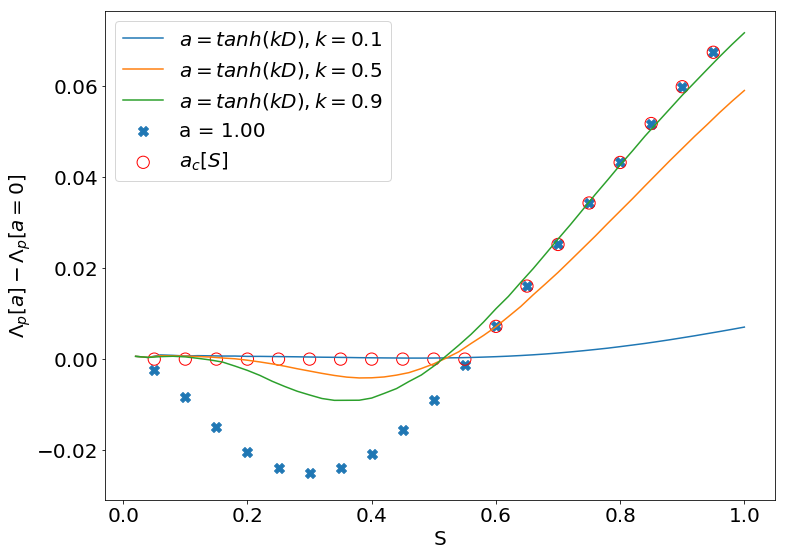}
\caption{The change of population growth rate ($\Lambda_p[a = \tanh(kD)] - \Lambda_p[a = 0]$) plotted against accumulation rate $S$. Here the key proteins are deleterious ($\lambda_0 = 1, \lambda_1 = 0, n = 2$). The blue cross show the values when $a$ is fixed at 1 ($\Lambda_p[a = 1] - \Lambda_p[a = 0]$) and the red circles show the change of the population growth rate when $a$ is equal to the optimal value $a_c$ at a given $S$.}
\label{adaptive_a}
\end{figure}

Here we replaced $D$ with $\sigma_b' +S$ since we assume the cell volume always is $1$ at birth and $2$ at division. In Figure ~\ref{adaptive_a}, we show the numerical value of population growth rate increase due to asymmetry ($\Lambda_p[a] - \Lambda_p [a = 0]$), while $a$ is a function of the amount of protein at division. Specifically, we set $a[D] = \tanh(kD)$ and tune $k$. In other words, we assume $a$ to increase from zero initially with the slope $k$ and asymptotically approach $a = 1$. Qualitatively, we expect the effective $a$ to be small when $S$ is small, and it becomes closer to $1$ as $S$ gets larger. We can check this intuition by writing down the equation for the moments of $\psi_{\rm birth}$ analogous to Equation \ref{moment}.

\begin{equation}
\hspace*{-2cm}
	\langle \Delta \sigma_b^n \rangle = \frac{1}{2^{n-1}} \sum_{m = 0}^{\lfloor n/2\rfloor} {n \choose 2m} \left<a[\Delta \sigma_b + 2S]^{2m} \Delta \sigma_b^{n - 2m} (\Delta \sigma_b + 2 S)^{2m} e^{-\Lambda_p \tau[\Delta \sigma_b + S]} \right>.
\end{equation}

If the slope of $a[D]$ is small compared to $a^2$, we can approximate $a[\Delta \sigma_b + 2S]^{2m}] \approx a[2S]$. Note that the population growth rate can be higher with an adaptive $a$ (e.g. with $k = 0.9$, $0.55 < S < 0.65$) than when $a$ is fixed at $a_c[S]$ (orange dots in Figure~\ref{adaptive_a}). Qualitatively, this suggests that there is an advantage if each cell tunes the asymmetry of segregation according to its level of damage.

\section{Discussion and outlook}
In this paper, we investigate how asymmetric segregation of key proteins in microbial organisms impact the population growth rate with a self-consistent equation (Equation \ref{self_consistent}) solved analytically and numerically. We adapt a physiological model proposed by \textit{Lin et al.} \cite{lin2019optimal} and apply the transport equation to set up an equation for the distribution of protein concentration at birth $\psi[\sigma_b]$ and population growth rate $\Lambda_p$ (Equation \ref{self_consistent}). The equation can be solved by updating the solutions iteratively as elaborated in Section \ref{iterate}. When the asymmetry is small ($a \ll 1$), $\Lambda_p$ can be approximated to $a^{2n}$ precision by solving a set of linear equations with respect to the moments of $\psi[\sigma_b]$ and $\Lambda_p$. Similar approximations can be made from the opposite limit, $a \approx 1$ (see \ref{a_near_1}). Changing the accumulation rate of key proteins, $S$, we confirm the sharp phase transition of $a_c$ for a deleterious protein and a smooth transition for a beneficial protein as \textit{Lin et al.} reported \cite{lin2019optimal}. In addition, since numerically solving Equation \ref{self_consistent} provides more precise prediction of $\Lambda_p$ than a direct simulation of an exponentially growing population, we are able to find that the phase transition in the case of damage is composed of a discrete jump to $a_c$ between 0 and 1 and a smooth increase to $a_c = 1$. We also explore different physiological models motivated by experimental studies. We discuss the case where the cell tunes the asymmetry with response to the amount of key proteins it has at division and how the adaptive strategy may minimize the disadvantage of asymmetry when the accumulation rate of damage is low. Finally, we show by generalizing our model that even if the segregation ratio is on average symmetric, the noise can have a similar effect as an asymmetric segregation. 

There are many other aspects of asymmetric division in biology where our theoretical framework might be useful for building a theoretical model \cite{levien2020non, moger2019microbial}. First, there are many other ways that microbes develop phenotypic diversity rather than due to asymmetric key protein segregation. For instance, budding yeast (\textit{Saccharomyces cerevisiae}) has cell size difference between a daughter (smaller cell) and mother cell (bigger cell), whereas when a \textit{Caulobacter crescentus} cell divides, one is non-motile and reproductive and the other swarms and does not divide until maturation \cite{curtis2010getting, obuchowski2008pfli}. In the former case, introduction of volume asymmetry turns out to create interesting correlations between the noise in growth and the population growth rate \cite{barber2020}. Second, mortality rate rather than volumetric growth rate may depend on the key protein concentration \cite{maisonneuve2008protein}. Lastly, connecting back to multicellular organisms, where the theory of aging (soma theory) was first proposed \cite{kirkwood1977evolution}, can be interesting both mathematically and biologically. Major difference from microbial system is that the cells interact with the neighbors and use the cues to make decisions \cite{moore2017creating}. Considering the geometry of the tissue might complicate the model beyond the level of a self-consistent equation can capture, but plants which have rigid tissue and simpler geometry may be a good starting point for a theoretical investigation \cite{pillitteri2016asymmetric}. 

\section{Acknowledgements and Funding Sources}
The authors thank the Amir group, the Desai lab, Andrew Murray, Jane Kondev and Doeke Hekstra for helpful discussions. J. M. acknowledges Yubo Su for helping improve numerical simulations. A. A. was supported by the NSF CAREER award number 1752024.

\appendix
\section{\label{simulation} Simulation details}
When we explicitly simulate the exponentially growing population, we start with 100 cells with uniform distribution of age. We run until the number of cells reaches $5 \times 10^6$ and find the slope of $ln(\sum V(t))$ after $t = 5$. This use of sum of volume rather than the number of cells for finding the population growth rate is because the volume increases continuously and the growth rate should be equivalent if the cell size is controlled. To solve Equation \ref{self_consistent}, we discretize the range of $\sigma_b$ and repeat the process described in section \ref{iterate} until the difference between 1 and $2 \int \int \psi_{\rm birth}^{(n)}[\sigma_b'] g(a') e^{-\Lambda_p^{(n-1)} \tau[\sigma_b']} da' d\sigma_b'$ is smaller than some tolerance value. This implies that $\Lambda_p^{(n-1)} \approx \Lambda_p^{(n)}$ and our solution for $\Lambda_p$ has converged. Python scripts for the simulations are on \href{https://github.com/jiseonmin/asymmetric_segregation}{our Github}.

\section{\label{a_near_1} $\Lambda_p$ when $a \approx 1$} 
The challenge for a highly asymmetric case is that the distribution $\psi_{\rm birth}[\sigma_b]$ is not symmetric around $S$. Because one of two newborn cells is rejuvenated at each division, we would expect the distribution to be skewed to the right and maximized near $\sigma_b = 0$. This means that we should expand $\psi_{\rm birth}[\sigma_b]$ around different values depending on when the cell has been rejuvenated for the last time (i.e. generational age), in order to solve the self-consistent equation. Categorizing cells based on their generational age, we can rewrite the self-consistent equation (\ref{self_consistent}) as,

\begin{eqnarray}
	\psi_0[\sigma_b] &= \int \sum_{i = 0}^\infty \psi_i[\sigma_b'] f(\sigma_b, \rm{R}| \sigma_b') e^{-\Lambda_p \tau[\sigma_b']} d\sigma_b',\nonumber \\
	\psi_i[\sigma_b] &= \int \psi_{i-1}[\sigma_b'] f(\sigma_b, \rm{S}| \sigma_b') e^{-\Lambda_p \tau[\sigma_b']} d\sigma_b (i > 0). \label{dist_near_1}
\end{eqnarray}

Here $\psi_i[\sigma_b]$ is the probability that a cell has protein concentration $\sigma_b$ at birth and the pole age $i$. The transition function for the rejuvenated cell is $f(\sigma, \rm{R}| \sigma_b') = \delta\left(\sigma - \frac{1-a}{2} (\sigma_b' + S) \right)$, and the senescent is $f(\sigma, \rm{S} | \sigma_b') = \delta\left( \sigma - \frac{1+a}{2}(\sigma_b' + S)\right)$. Note that the rejuvenating transition function does not depend on $\sigma_b'$ (parent cell) if $a = 1$. This is equivalent to the property used by Lin et al. \cite{lin2019optimal} to find a closed formula for $\Lambda_p[a = 1]$: there is a self-symmetry in a population tree in a stable exponential growth. 
\begin{eqnarray}	
	1 &= \sum_{i=0}^\infty \exp\left[- \sum_{j = 0}^i \Lambda_p \tau[jS] \right].
\end{eqnarray} 

Here, we can use the fact that every $\psi_i[\sigma_b]$ is a delta function if $a = 1$. 

\begin{eqnarray}
	\psi_i[\sigma_b] &= \int \psi_{i-1}[\sigma_b'] \delta(\sigma_b - (\sigma_b' + S) ) e^{-\Lambda_p \tau[\sigma_b']} d\sigma_b'\nonumber \\
	&= \psi_{i-1}[\sigma_b - S] e^{-\Lambda_p \tau[\sigma_b - S]} \nonumber \\
	&= \psi_0 [\sigma_b - iS] e^{-\sum_{j = 1}^i \Lambda_p\tau[\sigma_b - j S]} (i > 0)\\
	\psi_0[\sigma_b] &= \int \sum_{i = 0}^\infty \psi_i[\sigma_b'] \delta(\sigma_b) e^{-\Lambda_p \tau[\sigma_b']} d\sigma_b'\nonumber \\
	&=\delta(\sigma_b) \int  \sum_{i = 0}^\infty \psi_0[\sigma_b - iS] \exp\left[- \sum_{j = 0}^i \Lambda_p \tau[\sigma_b - jS] \right] d\sigma_b'
\end{eqnarray}

By integrating $\psi_0[\sigma_b]$ with respect to $\sigma_b$, we get
\begin{eqnarray}	
	1 &= \sum_{i=0}^\infty \exp\left[- \sum_{j = 0}^i \Lambda_p \tau[(i - j)S] \right] = \sum_{i=0}^\infty \exp\left[- \sum_{j = 0}^i \Lambda_p \tau[jS] \right],\label{self_similar_a_1}
\end{eqnarray} 

which is consistent to the previous paper \cite{lin2019optimal}. Let's call the unique solution for Equation \ref{self_similar_a_1}, $\Lambda_{p, 1}$. Near $a = 1$, we can find the approximate population growth rate by expanding $\Lambda_p = \Lambda_{p1} (1 + C_1 (1 - a) + C_2 (1-a)^2 + \cdots )$ where $a < 1$. As $a$ deviates from 1, we expect each delta function to broaden near $\sigma_b = i S$. Similar to the case of a small $a$, we will find the moments of $\psi_i[\sigma_b]$ starting from the lowest order. Here we will investigate the first order coefficient $C_1$. Importantly, finding the sign of $C_1$ will allow us to predict whether $\Lambda_p[a = 1]$ is a local minimum ($C_1 >0$) or a local maximum ($C_1 < 0$).

Let us define $\delta a = 1 - a$. To find $C_1$ we need to solve Equation \ref{dist_near_1} to $O(\delta a)$. Thus, we can rewrite the equations as
\begin{eqnarray}
	\psi_0[\sigma_b] &= \int \sum_{i = 0}^\infty \psi_i[\sigma_b']  \delta\left(\sigma_b - \frac{\delta a}{2} (\sigma_b' +S) \right)\nonumber\\
		& \exp\left[-\Lambda_{p1}\tau[iS](1 + C_1 \delta a) \left(1 + \frac{\tau'[iS]}{\tau[iS]} (\sigma_b' - iS) \right)  \right] d\sigma_b'\nonumber \\
	\psi_i[\sigma_b] &= \int \psi_{i-1}[\sigma_b'] \delta\left( \sigma_b - (1 - \delta a / 2) (\sigma_b' + S) \right) \nonumber \\
		&\exp\left[-\Lambda_{p1} \tau[(i-1)S] (1 + C_1 \delta a) \right. \nonumber \\
		&\left. \left(1 + \frac{\tau'[(i-1)S]}{\tau[(i-1)S]} (\sigma_b' - (i-1)S) \right) \right] d\sigma_b  
\end{eqnarray}
For brevity, from now on, we shall write $\Psi_i = \int \psi_i[\sigma_b] d\sigma_b$ and $\left<F[\delta \sigma_b]\right>_i = \int \psi_i [\sigma_b] F[\sigma_b - iS] d\sigma_b / \Psi_i$. First, let us integrate the $i$th equation $(i > 0)$,
\begin{eqnarray}
\hspace*{-2cm}
	&\Psi_i = \int \psi_{i - 1}[\sigma_b'] e^{-\Lambda_p \tau[\sigma_b']} d\sigma_b' \\
	&= e^{-\Lambda_{p1} \tau[(i-1)S]}  \int \psi_{i-1}[\sigma_b'] \nonumber \\
	    &\left(1 -\Lambda_{p1} \tau[(i-1)S] C_1 \delta a - \Lambda_{p1} \tau'[(i-1)S] ( \sigma_b' - (i - 1)S)\right)  d \sigma_b' \nonumber \\
		&=\Psi_{i-1} e^{-\Lambda_{p1} \tau[(i-1)S]} \left(1 -\Lambda_{p1} \tau[(i-1)S] C_1 \delta a
		-  \left<\delta \sigma_b\right>_{i-1} \Lambda_{p1} \tau'[(i-1)S]\right) \nonumber\\
		&=\Psi_0 e^{-\Lambda_{p1} \sum_{j = 0}^{i-1}\tau[jS]}
		 \left(1 -\Lambda_{p1} \sum_{k = 0}^{i-1}\tau[k S] C_1 \delta a -  \left<\delta \sigma_b\right>_k \Lambda_{p1} \tau'[k S]\right) + O(\delta a^2)\nonumber
\end{eqnarray}

For $i = 0$ we have
\begin{eqnarray}
	&\Psi_0 = \sum_{i = 0}^\infty \int \psi_i [\sigma_b'] e^{-\Lambda_p \tau[\sigma_b']} d\sigma_b' \nonumber \\
	&= \sum_{i = 1}^\infty \Psi_i \nonumber \\
	&=\Psi_0 \sum_{i = 0}^\infty e^{-\Lambda_{p1} \sum_{j = 0}^i\tau[jS]} \left(1 -\Lambda_{p1} \sum_{k = 0}^i\tau[k S] C_1 \delta a
		-  \left<\delta \sigma_b\right>_k \Lambda_{p1} \tau'[k S]\right) ,
\end{eqnarray}
to $O(\delta a)$ accuracy. 
Using the definition of $\Lambda_{p1}$ - $1 = \sum_{i = 0}^\infty e^{-\Lambda_{p1} \sum_{j = 0}^i \tau[jS]}$, we can extract an expression for the coefficient $C_1$.
\begin{eqnarray}\label{C1}
	C_1 &= -\frac{\sum_{i = 0}^\infty e^{-\Lambda_{p1} \sum_{j=0}^i \tau[jS]}  \sum_{k = 0}^i \tau'[kS] \langle \delta \sigma_b\rangle_k }{\sum_{i = 0}^\infty e^{-\Lambda_{p1} \sum_{j=0}^i \tau[jS]}  \sum_{k = 0}^i  \tau[kS] \delta a}\nonumber\\
		&= - \frac{\sum_{k = 0}^\infty \tau'[kS] \left<\delta \sigma_b\right>_k \sum_{i = k}^\infty e^{-\Lambda_{p1} \sum_{j = 0}^i \tau[jS]}}{\delta a \sum_{k = 0}^\infty \tau[kS] \sum_{i = k}^\infty e^{-\Lambda_{p1} \sum_{j = 0}^i \tau[jS]}}
\end{eqnarray}

Let us find $\left<\delta \sigma_b\right>_i$ for $i > 0$.

\begin{eqnarray}
	&\left<\sigma_b\right>_i = \frac{1}{\Psi_i} e^{-\Lambda_{p1} \tau[(i-1)S]}  \int \psi_{i-1}[\sigma_b']\nonumber \\
	&\left(\left(1 - \frac{\delta a}{2} \right)\left(\sigma_b' - (i-1)S + iS\right) - iS\right)  \nonumber\\
	& \left(1 -\Lambda_{p1} \tau[(i-1)S] C_1 \delta a - \Lambda_{p1} \tau'[(i-1)S] (\sigma_b' - (i - 1)S)\right) d\sigma_b' \nonumber\\
	&= - \frac{\delta a}{2} i S + \frac{1}{\Psi_i} e^{-\Lambda_{p1} \tau[(i-1)S]}  \Psi_{i-1} \left<\delta \sigma_b\right>_{i-1} \nonumber\\
	&= - \frac{\delta a}{2} \sum_{k = 0}^i (i - k)S \frac{\Psi_{i - k}}{\Psi_i} e^{-\Lambda_{p1} \sum_{j = 1}^k \tau[(i-j)]} \nonumber\\
	&+ \frac{\Psi_0}{\Psi_i} e^{-\Lambda_{p1} \sum_{j = 0}^{i-1} \tau[jS]} \left<\delta \sigma_b\right>_0\nonumber \\
	&= - \delta a \frac{S}{2}\frac{i(i+1)}{2} + \left<\delta \sigma_b\right>_0  + O(\delta a^2)
\end{eqnarray}

If $i = 0$,
\begin{eqnarray}
	&\left<\delta \sigma_b\right>_0 = \frac{1}{\Psi_0} \frac{\delta a}{2} \sum_{m = 0}^\infty \psi_m[\sigma_b']\left(\sigma_b' - mS + (m +1)S\right)\nonumber\\
		& \left[1 - \Lambda_{p1} \tau[mS] C_1 \delta a - \Lambda_{p1} \tau'[mS](\sigma_b' - mS)\right] d\sigma_b'\nonumber\\
		&= \frac{1}{\Psi_0} \frac{\delta a}{2} \sum_{m = 0}^\infty (m + 1)S \Psi_{m + 1}\\
		&= \frac{\delta a}{2} \sum_{m = 0}^\infty (m +1)S e^{-\Lambda_{p1} \sum_{n = 0}^m \tau[nS]} + O(\delta a^2)
\end{eqnarray}

This shows that $\left<\delta \sigma_b\right>_i$ is positive for small $i$ and becomes more and more negative as $i$ increases. Applying this to Equation \ref{C1}, we can find the slope near $a = 1$. 

\begin{equation}
\hspace*{-2cm}
	C_1 = \frac{S}{2} \frac{\sum_{i = 0 }^\infty e^{-\Lambda_{p1} \sum_{j = 0}^i \tau[jS]} \sum_{k = 0}^i \tau'[kS] \left( - \frac{k (k + 1)}{2} +\sum_{m = 0}^\infty (m + 1)e^{-\Lambda_p \sum_{n = 0}^m \tau[nS]} \right)}{\sum_{i = 0}^\infty e^{-\Lambda_{p1} \sum_{j = 0}^i \tau[jS]} \sum_{k = 0}^i \tau[kS]} 
\end{equation}

\section{\label{higher order} Finding $\Lambda_p[a]$ to $O(a^{2n}) (n > 1)$}
In order to calculate $\Lambda_p[a]$ to $O(a^{2n})$ accuracy, we write $\Lambda_p$ and $\tau[\sigma_b]$ as
\begin{eqnarray}
	\Lambda_p &= \frac{\ln 2}{\tau[S]} \left(1 + \sum_{k = 1}^n C_{2k} a^{2k}\right),\\
	\tau[\sigma_b'] &= \tau[S] \sum_{i = 0}^{2n} \frac{\tau^{(i)}[S]}{i! \tau[S]} \Delta \sigma_b'^i.
\end{eqnarray} 

Using these, we write
\begin{eqnarray}
	e^{-\Lambda_p \tau[\sigma_b]} &= \exp\left[-\ln 2 \left(1 + \sum_{k = 1}^n C_{2k} a^{2k}\right) \sum_{i = 0}^{2n} \frac{\tau^{(i)}[S]}{i! \tau[S]} \Delta \sigma_b^i \right]. 
\end{eqnarray}
Next, we should expand Equation \ref{moment} and truncate any terms with higher order than $a^{2n}$. We can also rewrite $\langle \sigma_b^m\rangle$  $(1 \leq m \leq 2n) $  as
\begin{eqnarray}
	\langle \Delta \sigma_b^{2k} \rangle &=  \sum_{j = k}^n D_{2k}^{(2j)} a^{2j},\\
	\langle \Delta \sigma_b^{2k - 1} \rangle &= \sum_{j = k}^n D_{2k + 1}^{(2j)} a^{2j}.
\end{eqnarray}
Maintaining only terms to $O(a^{2n})$ and matching the coefficients of $a^{2j} (1 \leq j \leq n)$, we can solve the system of equations for $D$ and $C$'s. We provide the SymPy code to find the coefficients on \href{https://github.com/jiseonmin/asymmetric_segregation}{our Github repository}.

\section{\label{distribution} The shape of $\psi_{\rm birth}[\sigma_b]$ changes depending on asymmetry $a$} 

Here we present some examples of the distribution of protein concentration at birth $\psi_{\rm birth}[\sigma_b]$, which we find by numerically solving Equation \ref{self_consistent} (red circles). In each simulation, we compare the numerically found distribution to a log-normal distribution and a normal distribution that have the same mean and variance as the numerical solution. In other words, we test whether it is possible to approximate the distribution of key protein concentration to a log-normal or a normal distribution and simplify Equation \ref{self_consistent}.

\begin{figure}[!h]
\centering
\subfloat[]{{\includegraphics[width = 7.5cm]{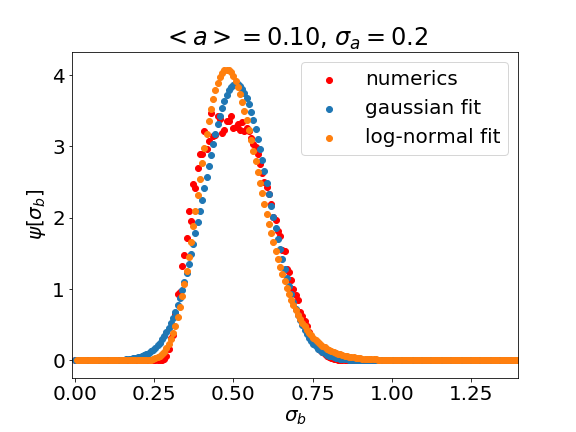} }}
\qquad
\subfloat[]{{\includegraphics[width = 7.5cm]{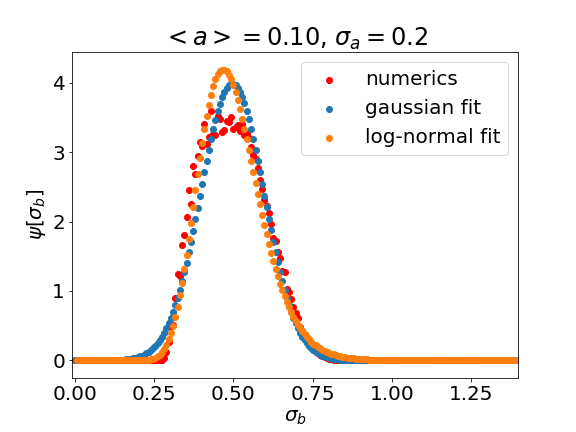} }}
\caption{Distribution $\psi_{\rm birth}[\sigma_b]$ can be approximated as a Gaussian or a log-normal distribution if $|\left<a\right>| \ll \sigma_a$). (a) Beneficial protein ($\lambda_0 = 0.2, \lambda_1 = 1.2, n = 2$), (b) Deleterious protein ($\lambda_0 = 1, \lambda_1 = 0, n = 2$). In both simulations, $\sigma_a = 0.2$ and $\left<a\right> = 0.1$}
\label{distribution1}
\end{figure}

\begin{figure}[!h]
\centering
\subfloat[]{{\includegraphics[width = 7.5cm]{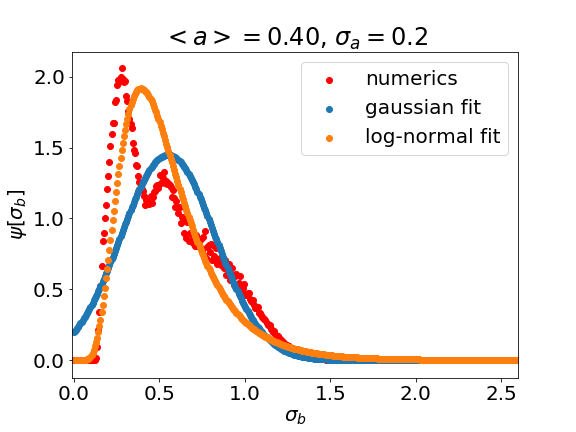} }}
\qquad
\subfloat[]{{\includegraphics[width = 7.5cm]{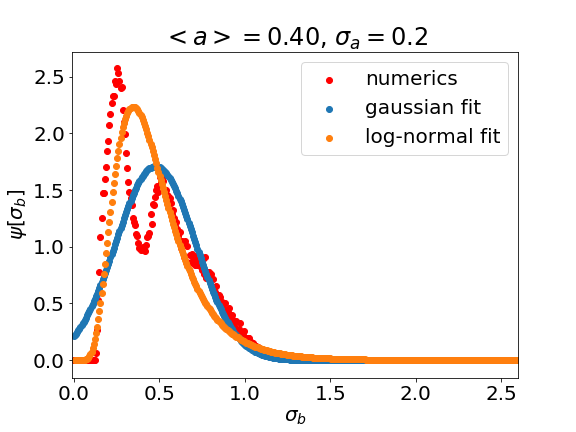} }}
\caption{Distribution $\psi_{\rm birth}[\sigma_b]$ is in general skewed and multi-modal. (a) Beneficial protein ($\lambda_0 = 0.2, \lambda_1 = 1.2, n = 2$), (b) Deleterious protein ($\lambda_0 = 1, \lambda_1 = 0, n = 2$). In both simulations, $\sigma_a = 0.2$ and $\left<a\right> = 0.4$}
\label{distribution2}
\end{figure}

When $a$ is small and the segregation is noisy ($\sigma_a > 0$), the distribution can be approximated as a Gaussian distribution or a log-normal distribution. However, as $a$ increases, the skewness of the distribution grows and the distribution becomes multi-modal. This implies that with small average asymmetry, the first two moments of the distribution have sufficient information for approximately solving Equation \ref{self_consistent} for $\Lambda_p$, but as the segregation becomes more and more asymmetric, higher order terms of $a$ are necessary to accurately calculate the population growth rate. The exact value of $a$ where the Gaussian approximation breaks down depends on the shape of $\lambda[\sigma]$ and $S$. 

\begin{figure}[!h]
\centering
\subfloat[]{{\includegraphics[width = 7.5cm]{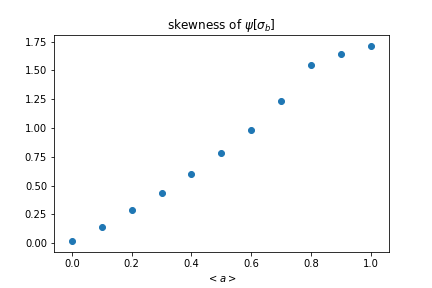} }}
\qquad
\subfloat[]{{\includegraphics[width = 7.5cm]{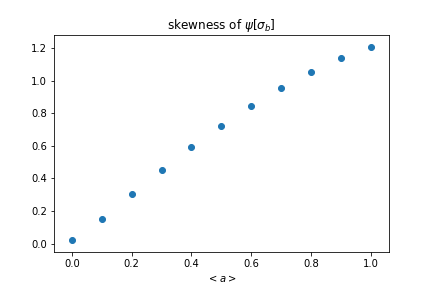} }}
\caption{Gaussian approximation only holds for the small asymmetry case because the skewness of distribution grows as a function of $a$. To illustrate this effect, we plot the skewness of $\psi_{\rm birth}{\sigma_b}$ against the average of $a$ for the damage and benefit case. To visualize the change of distribution, we show distribution with different values of a. The distributions are found numerically from the iterative solution of the self-consistent equations.  (a) Beneficial protein ($\lambda_0 = 0.2, \lambda_1 = 1.2, n = 2$), (b) Deleterious protein ($\lambda_0 = 1, \lambda_1 = 0, n = 2$). In both simulations, there is very small noise in $a$ ($\sigma_a = 0.02$). In both simulations, we use $S = 0.5$ and assume deterministic segregation ($\sigma_a = 0$).}
\label{skewness}
\end{figure}

\clearpage

\bibliography{jminJSTAT_resubmission}

\end{document}